\documentstyle[11pt,aaspp4]{article}

\def \m{\ifmmode M_\odot\else M$_\odot$\fi}

\def\deg{\ifmmode^\circ\else$^\circ$\fi} 
\slugcomment {Accepted for Publication in Astrophysical Journal Letters}

\begin{document}

\title{NICMOS OBSERVATIONS OF THE PRE-MAIN-SEQUENCE 
PLANETARY DEBRIS SYSTEM HD~98800\footnote{Based on 
observations with the NASA/ESA Hubble Space Telescope 
obtained at the Space Telescope Science Institute, which is 
operated by Association of Universities for Research in 
Astronomy, Incorporated, under NASA contract NAS5-26555.}}

\author{Frank J. Low, Dean C. 
Hines \& Glenn Schneider}
\affil{Steward Observatory, The University of Arizona, 933 
N. Cherry Ave., Tucson, AZ 85721} 
\affil{flow, dhines, gschneid@as.arizona.edu} 
\authoraddr{Steward Observatory, The University of Arizona, 
Tucson, AZ 85721}



\begin{abstract}
    
Spectral energy distributions (SEDs) from 0.4 to 4.7$\mu$m 
are presented for the two principal stellar components of 
HD~98800, A and B. The third major component, an extensive 
planetary debris system (PDS), emits $> 20\%$ of the 
luminosity of star B in a 
blackbody SED at $164\pm 5K$ extending from mid-IR to 
millimeter-wavelengths.  At 0.95$\mu$m a preliminary upper 
limit of $< 0.06$ is obtained for the ratio of reflected 
light to the total from star B. This result limits the 
albedo of the PDS to $< 0.3$.  Values are presented for 
the temperature, luminosity, and radius of each major 
systemic component.  Remarkable similarities are found 
between the PDS and the interplanetary debris system around 
the Sun as it could have appeared a few million years after 
its formation.

\end{abstract}
 
 \vfill \eject

\section{INTRODUCTION}
Long known as a visual double with less than 1\arcsec\ 
separation, HD~98800 (SAO 179815; IRAS P11195-2430) was 
found by {\it IRAS} to contain the brightest planetary 
debris system (PDS) in the sky.  Now we find that in terms 
of dimensions, configuration, temperature, and likely 
origin, this very young PDS bears remarkable similarity to 
the zodiacal dust bands in our solar system formed and 
maintained for four billion years by the asteroid families, 
a phenomenon also discovered by {\it IRAS} (Low et al.  
1984).  Recently identified as one of 11 known members of 
the nearby TW Hydrae Association (Kastner et al.  1997; Webb 
et al.  1999), the HD~98800 (TWA~4) system is comprised of 
two similar K dwarfs that have not yet reached the main 
sequence.  Torres et al.  (1995) reported that both stars 
are spectroscopic binaries with periods of 262 (Aa+Ab) and 
315 days (Ba+Bb).  At a distance of $46.7\pm 6$pc, measured 
by the {\it Hipparcos} satellite, the two brightest 
components of HD~98800 are well resolved by {\it HST} from 
0.4 to 2$\mu$m.  See Soderblom et al.  (1998) for BVI 
photometry of the two major components using WFPC2 on HST, 
and for estimates of their masses ($\sim$1 M$_{\odot}$) and 
age ($<$10 Myr).  From the ground at 4.7 and 9.8$\mu$m, 
Gehrz et al.  (1999) showed that the visual companion 
(component B) is actually at the center of the PDS with only 
a small amount of dust possible around star A. Star B lies 
north of star A by about 0\farcs8.  Separations of Ab from 
Aa and of Bb from Ba are of order 1 AU ($0\farcs02$), and 
the luminosity of Bb is thought to be greater than that of 
Ab.

Using the sub-arcsecond resolution and dynamic range in 
excess of one million afforded by the Near Infrared Camera 
and Multi-Object Spectrometer (NICMOS) onboard {\it HST}, we 
have resolved the two primary components in five bands from 
0.95 to 1.9$\mu$m, spanning the peak of their spectral 
energy distributions.  Our objective was to obtain precise 
relative and absolute photometry of the two stellar 
components and to search for a halo of scattered or 
reflected light from the PDS, realizing that all other 
resolved planetary debris systems scatter and emit about 
equally.  However, based on our predicted inner diameter of 
$< 10$~AU and small cross-sectional area for the PDS, we 
cannot be confident of separating the reflected light from 
direct starlight.

\section{NICMOS OBSERVATIONS AND REDUCTIONS}
Table 1 summarizes key observational parameters of the two 
{\it HST} orbits (GTO 7232) devoted to these measurements.  
Spanning 306 days, the two orbits were aimed at obtaining 
high fidelity ``sub-stepped'' images using four narrow band 
filters and one centrally located medium band filter.  Four 
images were planned at each epoch and at each wavelength 
(Table 1) using a spiral dither pattern (Mackenty et al.  
1997) to allow for bad pixel replacement and to improve the 
spatial sampling (from $\sim 2$ AU to $\sim 1$ AU).  
Unfortunately, in the first orbit only half of the images 
were obtained due to a pointing error.  Care was taken to 
correct for effects of non-linear response in the detectors, 
including
persistence, for cosmic ray events, and for flat fielding of 
the images.  In-flight flat-field images and the HD~98800 
raw images were reduced with in-flight dark frames.

Table 2 summarizes flux densities and their ratios derived 
from the complete set of 18 images.  Our absolute 
calibrations of the final background subtracted images are 
based on preliminary photometric calibrations kindly 
provided by M. Rieke (1999) prior to their publication.  
Flux densities for magnitude zero stars, accurate to $\pm 
3\%$, are listed in Table 2, column 6, to facilitate 
conversion to magnitudes and for comparisons with other 
calibrations.
When the images are combined, peak signal-to-noise is high 
(S/N $> 500$), and flat-fields good only to $0.3\%$ dominate 
the relative photometric errors, while uncertainties in the 
absolute flux densities are dominated by calibration errors.  
For completeness we have included the flux densities derived 
from the B, V and ``I'' magnitudes reported by Soderblom et 
al.  (1998), and the ratios (B/A) from our reprocessing of 
their publicly available images.

Precise NICMOS ratios were determined from each calibrated 
image using the software package IDP3 (Lytle et al.  1999) 
as follows: a scaled copy of the image was shifted to center 
star A precisely over star B and the two images were 
subtracted after varying the scale factor to minimize the 
residual.  Then the process was repeated in reverse order 
and, finally, the image of star B was divided by that of 
star A. The residuals were used to determine the error.  
Subtraction of a PSF-star gave consistent, but noisier 
results.  A similar procedure was used to derive the listed 
ratios from the WFPC2 images.

\section{DISCUSSION}

Figure (1) shows the spectral energy distributions of 
components A and B based on WFPC2 and NICMOS data and the 
results from Gehrz et al.  (1999).  For the PDS we include 
color-corrected Faint Source Catalog flux densities from 
{\it IRAS} and measurements at sub-mm wavelengths (see Table 
2).  Shown in Figure (1a) are blackbody fits to each stellar 
component and to the PDS. Figure (1b) shows the percentage 
deviations of the observations of the two stars from the 
Planck Law.  The corresponding blackbody temperatures 
(T$_{\rm BB}$) for the two primary stars and for the PDS are 
included in Table 3, along with errors estimated from the 
fitting process and from the flux density errors.  For 
comparison, we list effective temperatures for the stars 
based on the B-V colors from Soderblom et al.  (1998) using 
the Bessell (1979) color-temperature relation, but treating 
both A and B as single stars.  Also, we have attempted to 
fit these data using model SEDs based on ground-based 
photometry of K dwarfs of the same spectral types.  
Unexpectedly, these unpublished models of main sequence 
K-stars (M. Meyer 1999) do not represent the measured flux 
densities of A and B short of 2$\mu$m as well as do the 
simple blackbody models.  However, beyond 2$\mu$m the SEDs 
of these stars appear to drop below the simple blackbody 
model.  The luminosities listed in Table 3, derived by 
integration of the blackbody models, may be high by as much 
as 3$\%$ due to this effect.  Because the B-V colors are 
close to normal for stars with spectral types K5 and K7, we 
do not include an interstellar extinction correction.  
Neglecting the uncertainty of the distance, we believe the 
errors of the luminosity determinations are of order 
$\pm5\%$.  Taking our measured luminosities and treating our 
blackbody temperatures as good approximations to the true 
effective temperatures, current PMS 
models\footnote[1]{http://www-astro.phast.umass.edu/data/tracks.html} 
give plausible values for mass, but are in conflict with 
other age indicators.

Using the luminosity and temperature of components A and B 
their radii were calculated.  In neglecting their duplicity 
we note that this oversimplification is more significant for 
B than for A (see Soderblom et al.  1996).  The values in 
Table 3 show that the cooler star, B, is slightly larger 
than star A, a result well within the precision of the 
relative measurements involved.  Pre-main sequence stars may 
well have this property.  However, when both components of 
star B are considered, they each can have radii smaller than 
that of star Aa (Ab is almost negligible), and as previously 
noted, some portion of the power emitted by Ba and Bb will 
be reflected by the PDS adding to the star's apparent 
luminosity.  Figure 1 shows that there are slight 
differences between the two primary components when compared 
in detail relative to their respective blackbodies.  Also, 
star A varies at all wavelengths by as much as a few percent 
and significant coronal activity has been reported (e.g. 
Fekel \& Bopp 1993; Henry \& Hall 1994).
 
At 0.95$\mu$m, where our resolution is highest, we have been 
unable to detect starlight reflected by the PDS. From the 
residuals of the flux-scaled PSF subtraction, we place a 
tentative upper limit of $< 6\%$ for the reflected light 
relative to the direct emission from Ba+Bb.  These residuals 
have been compared to quantitative predictions based on a 
simplified geometrical model of the IR source convolved with 
star A (which serves as the PSF).  From Table 3 we see that 
the power emitted in our direction by the PDS divided by the 
total power emitted by the Ba+Bb systemic components and 
reflected from the PDS is 0.19.  Therefore, at 0.95$\mu$m we 
have placed an upper limit on the albedo of the reflecting 
material of 0.3, well within the range expected for such 
material.  Asteroids have albedos that are generally lower 
than 0.3 [see results from {\it IRAS} reported by Tedesco et 
al.  (1989)].  Note also that we have direct observational 
evidence from the work of Skinner, Barlow \& Justtanont 
(1992) that the emitting material contains silicates.  
However, in our filters we see no indication of coloration 
of the stellar emission from component B as might be caused 
by reflection from the PDS. This suggests that the albedo is 
low at all bands from 0.4 to 2$\mu$m.

Next, we must explain how a circumstellar envelope can form 
and be maintained over millions of years with geometry, 
opacity, and stability sufficient to produce the system that 
we observe.  When Zuckerman \& Becklin (1993) reported the 
extraordinarily high value of L$_{\rm IR}$/L$_{\rm bol}$ for 
HD~98800, they pointed out that the lifetime of a 
circumstellar dust cloud thick enough to absorb 15$\%$ or 
more of the stellar emission would quickly evolve into a 
thin disk.  Consider the case of a failed terrestrial planet 
of sufficient mass to form dense ``dust bands'' which 
overlap and fill in a belt subtending 20$\%$ of the 
celestial sphere around the stars.  The required range of 
orbital inclinations is -12\deg\ to $+12\deg$.  In the solar 
system the dust bands lie 10\deg\ above and below the 
zodiacal plane, and are constantly replenished by 
collisional erosion of asteroids in those same orbits.  The 
much larger quantity of dust in the HD~98800B planetary 
debris system follows from its recent formation

Just as we calculated the radii of the stars, we have also 
calculated an ``equivalent'' radius for the material 
orbiting around star B, even though in projection the IR 
source, clearly, is not circular in shape.  The result 
listed in Table 3, 2 AU, is consistent with our proposed 
geometry of the PDS, and constrains that portion of the belt 
that is visible directly.
The actual radii of the inner debris orbits, about 4.5 AU, 
can be predicted because the material must be in radiative 
equilibrium with the measured emission of the two central 
stars.  In the solar system, where the corresponding 
temperatures are 200 to 250 K, the dust is optically thin, 
of order $10^{-8}$, and its emission follows a blackbody at 
least out to 100$\mu$m.

In HD~98800B the dust opacity remains high from 7 to at 
least 1000$\mu$m, but we find no indication of attenuation 
of light from star B. This implies that the orbital plane of 
the dust system is inclined to our line of sight by more 
than 12\deg.  The most likely value is near 45\deg\ since 
the emitting area must equal that of the equivalent disk, 
$\sim$ 12 sq.  AU. Because most of the 165 K emitting 
surface is obscured by colder particles in larger orbits, 
the actual ratio of IR to optical luminosity is probably 
higher than the observed ratio of 0.19 reported here.
Clearly, the IR emission is not isotropic.  Using the 
predicted 4.5 AU inner radius of the belt, an equivalent 
thickness of 1 mm, and terrestrial density, a rough estimate 
of minimum mass is 0.6 M$_{\oplus}$.

In conclusion, we now have enough information about the 
stars in HD~98800 to test, and perhaps improve, models of 
PMS dwarfs.  We can also construct plausible models of 
recently formed planetary debris systems with interesting 
similarities to our own
planetary system.  We note that as yet no other examples of 
very young systems of this type have been found, indicating 
that the lifetime of this optically thick phase is probably 
rather short.  Future observations using adaptive optics on 
the largest telescopes and space missions such as NGST 
should resolve both the thermal and reflected components of 
the PDS around HD~98800B. Infrared surveys now in progress 
and in the planning stage with SIRTF should provide new and 
better means of locating planetary debris systems, thus
providing an answer to the question of how often terrestrial 
planets are formed.

\acknowledgments { Valuable assistance from our colleagues, 
M. Meyer, M. Rieke, D. McCarthy and E. Becklin is much 
appreciated, and we thank A. Shultz, D. Golombek, P. 
Stanley, and I. Dashevsky for their help at STScI. This work 
is supported by NASA grant NAG5-3042 to the NICMOS 
instrument definition team.  }

\newpage

\begin{figure}[1]
\includegraphics{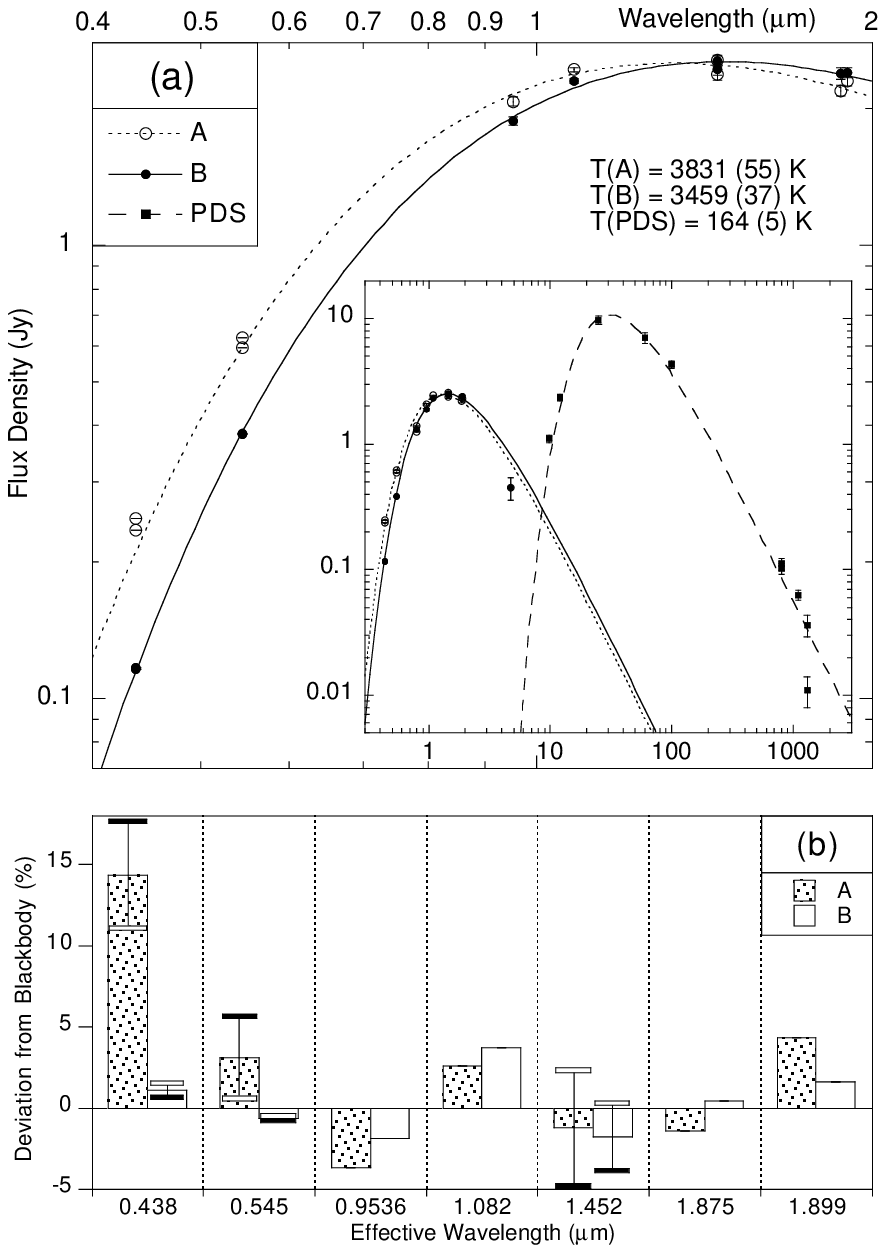}
\vspace{11truecm}
\caption{(a) SEDs of primary components with 
blackbody fits.  Expanded view shows the measurements used 
for the fits to A \& B. (b) Spectral and temporal deviations 
of mean flux densities from blackbody fits.  First epoch 
indicated by black (see Table 2).}
\end{figure}


\begin{deluxetable}{lllrcccc}
\tablenum{1}
\tablewidth{0pt}
\tablecaption{NICMOS Observation Log}
\tablehead{\colhead{Date} & \colhead{Filter} & 
\colhead{Mode} & \colhead{Exposure}
& \colhead{Dither} & 
\colhead{$\lambda/2D$\tablenotemark{a}} & \colhead{Frames} \\
\colhead{(UT)} & \colhead{} & \colhead{} & \colhead{(sec)}
& \colhead{(pixels)} &  
\colhead{($\arcsec$)} & \colhead{}} 
\startdata
1997 Jul 3&F108N&STEP2	&67.78	&50.5	&0.0465	&2\nl
1997 Jul 3&F145M&SCAMRR	&7.30	&50.5	&0.0624	&2\nl
1997 Jul 3&F187N&STEP2	&59.82	&50.5	&0.0806	&2\nl
1998 May 1&F095N&STEP2	&135.56	&30.5	&0.0410	&4\nl
1998 May 1&F145M&SCAMRR	&16.24	&30.5	&0.0624	&4\nl
1998 May 1&F190N&STEP2	&135.56	&30.5	&0.0816	&4\nl

\tablenotetext{a}{NIC1 pixel scale = $0.0432\arcsec \times 
0.0432\arcsec$, 1.0 AU = $0.021\arcsec$}
\enddata
\end{deluxetable}

\begin{figure}
\includegraphics{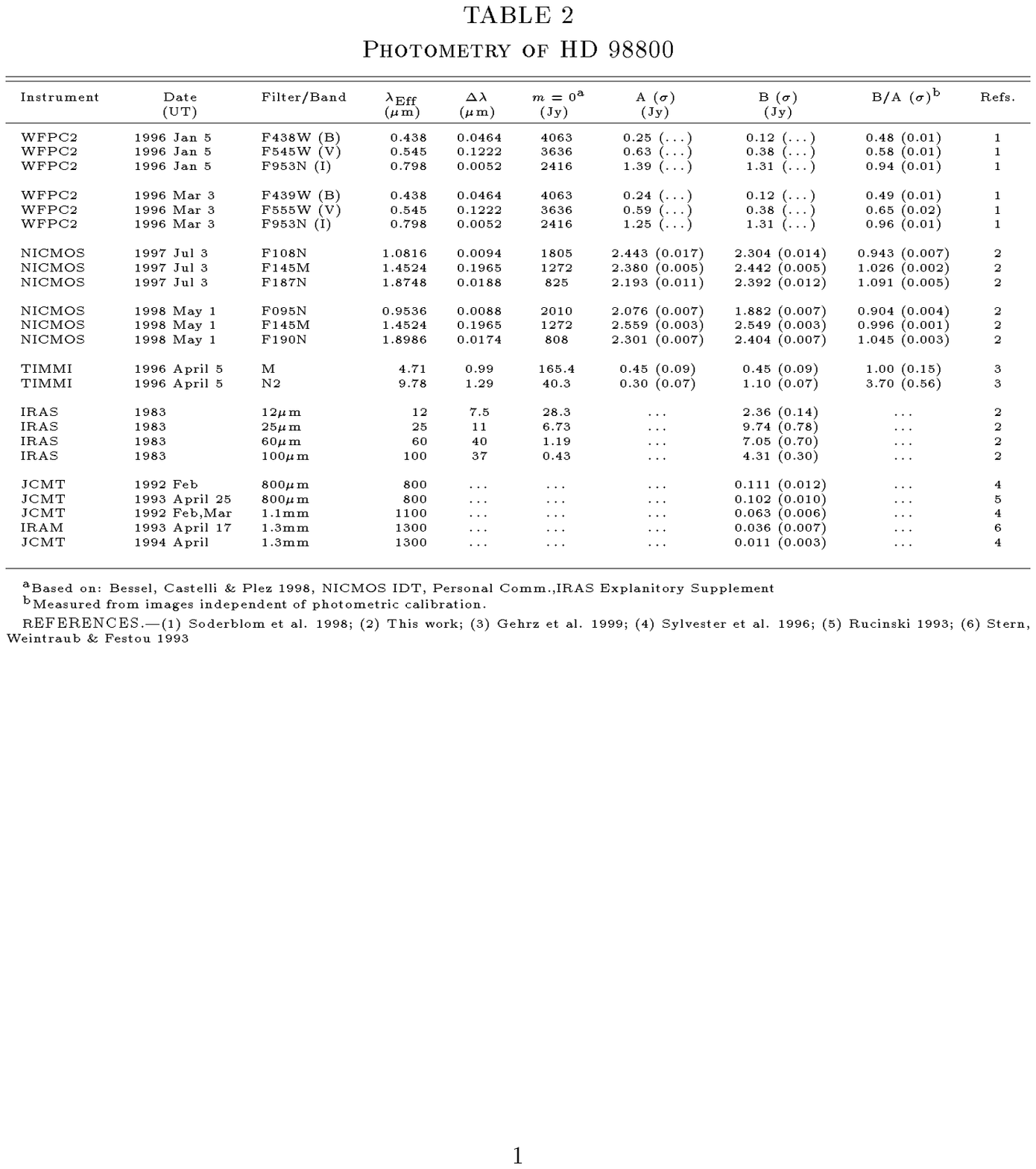}
\vspace{11truecm}
\end{figure}

\begin{deluxetable}{lcccccc}
\tablenum{3}
\tablewidth{0pt}
\tablecaption{Derived properties of HD~98800}
\tablehead{\colhead{Comp.} & 
\colhead{Sp.\tablenotemark{a}} & \colhead{T$_{\rm 
eff}$(B-V)\tablenotemark{b}} & \colhead{T$_{\rm BB}$} & 
\colhead{L\tablenotemark{c}} & \colhead{R\tablenotemark{d}}
&\colhead{Mass} \\
\colhead{} & \colhead{} & \colhead{(K)}
& \colhead{(K)} & \colhead{(L$_{\odot}$)} 
&\colhead{(R$_{\odot}$)} & \colhead{(M$_{\oplus}$)}}
\startdata
A&K5	&4100	&3831 (55)&0.63 (0.03)	&1.94 (0.06) &\ldots\\
B&K7	&3500	&3459 (37)&0.57 (0.03)	&2.13 (0.06) &\ldots \\
& &	&	&	&	& 	 \\
PDS&\ldots&\ldots&164 (5)& 0.11 (0.01)	&417 (34) & 0.6	\\
PDS&\ldots&\ldots&\ldots&0.19L$_{\rm StarB}$&1.94 AU &\ldots \\

\tablenotetext{a}{Based on B-V colors from Soderblom et al.  
(1998)}

\tablenotetext{b}{Effective temperature based on B-V 
colors.}

\tablenotetext{c}{Luminosity assuming isotropy and d = 46.7 
pc (Fig.  1)}

\tablenotetext{d}{The effective/equivalent radius of a blackbody of 
temperature T$_{\rm BB}$.  }
\enddata
\end{deluxetable}

\end{document}